\documentclass[graybox]{svmult}

\usepackage{mathptmx}       
\usepackage{helvet}         
\usepackage{courier}        
\usepackage{type1cm}        
\usepackage{makeidx}         
\usepackage{graphicx}        
\usepackage{multicol}        
\usepackage{url}        
\usepackage[bottom]{footmisc}

\makeindex             

\begin{document}

\title*{Applying the Background-Source separation algorithm to Chandra Deep Field South data} 
\titlerunning{Applying the BSS algorithm}
\author{F. Guglielmetti$^{1}$ \and H. B\"ohringer$^{1}$ \and R. Fischer$^{2}$ \and P. Rosati$^{3}$ \and P. Tozzi$^{4}$}
\institute{Fabrizia Guglielmetti \and Hans B\"ohringer
\at Max-Planck-Institut f\"ur extraterrestrische Physik, Giessenbachstrasse, D-85748 Garching, Germany, \email{fabrizia@mpe.mpg.de, hxb@mpe.mpg.de}
\and Rainer Fischer \at Max-Planck-Institute f\"ur Plasmaphysik, Boltzmannstrasse 2, D-85748 Garching, Germany, \email{Rainer.Fischer@ipp.mpg.de} \and Rosati Piero \at European Southern Observatory, Karl-Schwarzschild-Strasse 2, D-85748 Garching, Germany, \email{prosati@eso.org} \and Paolo Tozzi \at INAF-OATs, Via Tiepolo 11, I-34143 Trieste, Italy, \email{tozzi@oats.inaf.it}}
\maketitle
\abstract*{A probabilistic two-component mixture model allows one to separate the diffuse background 
from the celestial sources within a one-step algorithm without data censoring. The background is 
modelled with a thin-plate spline combined with the satellite's exposure time. 
Source probability maps are created in a multi-resolution analysis for revealing faint and extended 
sources. 
All detected sources are automatically parametrized to produce a list of source positions, fluxes and 
morphological parameters. The present analysis is applied to the \emph{Chandra} Deep Field South 2\ Ms 
public released data. Within its 1.884\ ks of exposure time and its angular resolution (0.984\ arcsec), 
the \emph{Chandra} Deep Field South data are particularly suited for testing the Background-Source 
separation algorithm.
} 

An analysis is performed to test the sensitivity and the internal consistency of the Background-Source separation (BSS) 
algorithm (see Ref.~\cite{gugli:2009} and proceeding Guglielmetti et al.~in this volume)  
with sources on real fields from pointed observations. 
The employed field is the \emph{Chandra} Deep Field South (CDF-S) 2\ Ms data \cite{luo:2008}. 
The optimal energy band to detect the emission from both point-like and extended sources is between  
0.5-2.0\ keV. Therefore, this test is concentrated on this energy range.  
The main advantages of testing real data with respect to simulated ones reside on the fact that 
real data are characterized by a complex background, a complex point-spread function dependence 
across the field, source confusion and a wide range of source properties. 
These characteristics intrinsic to real observations are not easily 
elaborated with artificial data. 
Therefore, the CDF-S 2\ Ms data are separated in four images of 500\ ks exposure time each. 

The BSS algorithm is applied on each of the four images. The exponential prior probability density function 
of the source signal is chosen and $25$\ pivots equally spaced are used for the
background rate estimation. 
Scales in the range value $0.5-13$\ arcsec are used in the multiresolution
analysis. 
A threshold value of $P_{\rm source}\ge0.9$ is chosen to separate
false-positives in source detection from true sources. 
No contaminations due to steep changes in the exposure time map are 
seen both in the background map and in the source probability maps. 
The multiresolution analysis provides for the detection 
of a wide range of source fluxes and their complex morphologies.

The internal consistency of the BSS algorithm is tested comparing each CDF-S 500\ ks data in pairs. 
Source positions (right ascension and declination), fluxes and extent (i.e.~the estimated size of the 
detected sources) are taken into account. 
The difference of position estimates are within $1\sigma$ (Fig.1 {\bf a}), while the ones of 
fluxes and extents are within $3\sigma$ (Fig.1 {\bf b}). 
Although $70\%$ of the sources in the CDF-S region are characterized by $X$-ray variability 
\cite{paolillo:2004},
Poisson fluctuations and contaminations by other sources in the fields  
can increase the uncertainties estimated for the source flux and extent measurements.
The BSS estimates of source parameters are internally consistent. 

A sensitivity analysis is performed on the four CDF-S 500\ ks data and the results are compared to 
published ones: CDF-S 1\ Ms \cite{giacconi:2002} and 2\ Ms \cite{luo:2008} data.  
The information about the sensitivity and the reliability of the survey are described by the sky coverage and the 
log\emph{N}-log\emph{S} distribution. 
The estimated sky coverage and the log\emph{N}-log\emph{S} distribution depend on the algorithm 
employed for source detection. 
The BSS background maps are used to construct the flux limit map of each estimated  
sky coverage. Hence, vignetting effects and background variations are already accounted in the 
coverage. 
The log\emph{N}-log\emph{S} distributions are computed from the respective sky coverage. 
It results that the log\emph{N}-log\emph{S} distributions obtained with the four CDF-S 500\ ks data are in 
agreement with the published ones in Refs.~\cite{giacconi:2002,luo:2008}: See Fig.1 {\bf c}.  

Applying the BSS algorithm to the CDF-S data, we prove that the BSS algorithm provides for a 
reliable detection of sources and estimation of source and background parameters. An extensive application of the technique is addressed in a forthcoming paper (Guglielmetti et al., in preparation).
\begin{figure}[ht]
 \begin{center}
\begin{minipage}[c]{0.6\linewidth}
\hspace{-2.6cm}
 \includegraphics[width=.65\linewidth,angle=90]{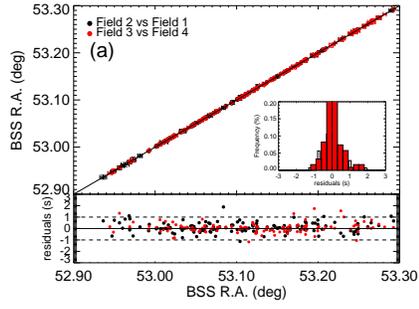}\includegraphics[width=.65\linewidth,angle=90]{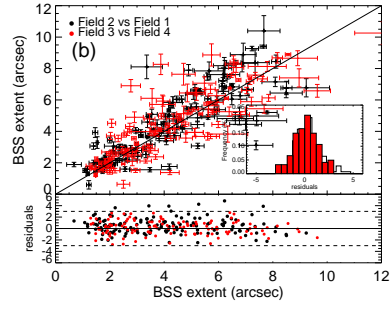}\hspace{\fill}
\end{minipage}
 \begin{minipage}[t]{0.5\linewidth}
\hspace{.1cm}
   \includegraphics[width=.7\linewidth,angle=90]{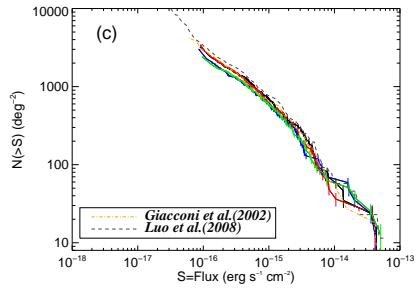}
  \end{minipage}\hfill
 \begin{minipage}[t]{0.45\linewidth}
  \label{Fig1}
\vspace{-4.cm}
\caption{Internal consistency and sensitivity analyses. {\bf a}: source position (right ascension). {\bf b}: source extent. {\bf c}: log\emph{N}-log\emph{S} distribution. In panels {\bf a-b}, field 1-2-3-4 indicate the four analysed CDF-S 500\ ks images. In panel {\bf c}, the log\emph{N}-log\emph{S} distributions obtained from the analysis of the four CDF-S 500\ ks images are drawn with a continuous line in red, green, yellow and black. }
 \end{minipage}
 \end{center}
\end{figure}
%
%
%
%
%
%

\end{document}